\documentclass[a4paper,aps,floatfix,twocolumn,pre]{revtex4-1}

\usepackage{docmute}

\usepackage{amsfonts,amssymb,amsmath,amsthm,bm}
\usepackage{graphicx}

\makeatletter
\let\MYcaption\@makecaption
\makeatother

\usepackage{caption}
\makeatletter
\let\@makecaption\MYcaption
\makeatother

\usepackage{subcaption}
\usepackage[colorlinks]{hyperref}
\usepackage{color}

\begin{document}
	\title{Effects of Boundary Conditions on Magnetic Friction}%
	
	\author{Kentaro Sugimoto}
	\email[]{tarotene@iis.u-tokyo.ac.jp}
	\affiliation{Department of Physics, The University of Tokyo, 5-1-5 Kashiwanoha, Kashiwa, Chiba 277-8574, Japan}
	\affiliation{RIKEN Cluster for Pioneering Research (CPR), 2-1 Hirosawa, Wako, Saitama 351-0198, Japan}
	
	
	\begin{abstract}
		We consider magnetic friction between two square lattices of the ferromagnetic Ising model of finite thickness. We analyze the dependence on the boundary conditions and the sample thickness. Monte Carlo results indicate that the setup enables us to control the frictional force by magnetic fields on the boundaries. In addition, we confirm that the temperature derivative of the frictional force as well as that of the boundary energy has singularity at a velocity-dependent critical temperature.
	\end{abstract}

	\maketitle
	
	\section{Introduction}\label{sec:Introduction}
	
	Many studies have revealed the nature of friction and its applications \cite{Persson2000,Yoshino2007,Saito2007,Goryo2007,Inui2007,Zaloj1999,Sasaki2007,Filippov2008,Dudko2002,Braiman2003,Tshiprut2005,Tshiprut2009,Guerra2008} and yet the friction has been difficult to understand from a microscopic point of view. In non-equilibrium statistical physics, it is also an unanswered question how macroscopic sliding motion of objects makes their microscopic quasi-particle excitations on the sliding surface and how the excited energy dissipates. Several experimental facts suggest that physical degrees of freedom, such as phonon \cite{Highland2006,Coffey2005,Torres2006,Dag2004,Liebsch1999,Tomassone1999,Persson1999}, orbital motion of electrons \cite{Sokoloff2018,Conache2010,Park2007,Persson1999} and magnetic moment of spins \cite{Wolter2012,Ouazi2014}, play roles of dissipation channels. Especially for the magnetic moment, Monte Carlo simulations of classical spin systems by the use of the Monte Carlo simulations and the analysis based on the Landau-Lifshitz-Gilbert equation \cite{Kadau2008,Magiera2009,Hucht2009,Magiera2011,Hilhorst2011,Igloi2011,Heinrich2012,Hucht2012,Li2012,Angst2012,Magiera2013,Li2016} have revealed several facts regarding the friction due to magnetism from the viewpoints of statistical mechanics.
	
\begin{figure}[h]
		\begin{subfigure}[b]{.49\linewidth}
			\centering
			\includegraphics[width=0.75\linewidth,clip,bb=0 0 205 284]{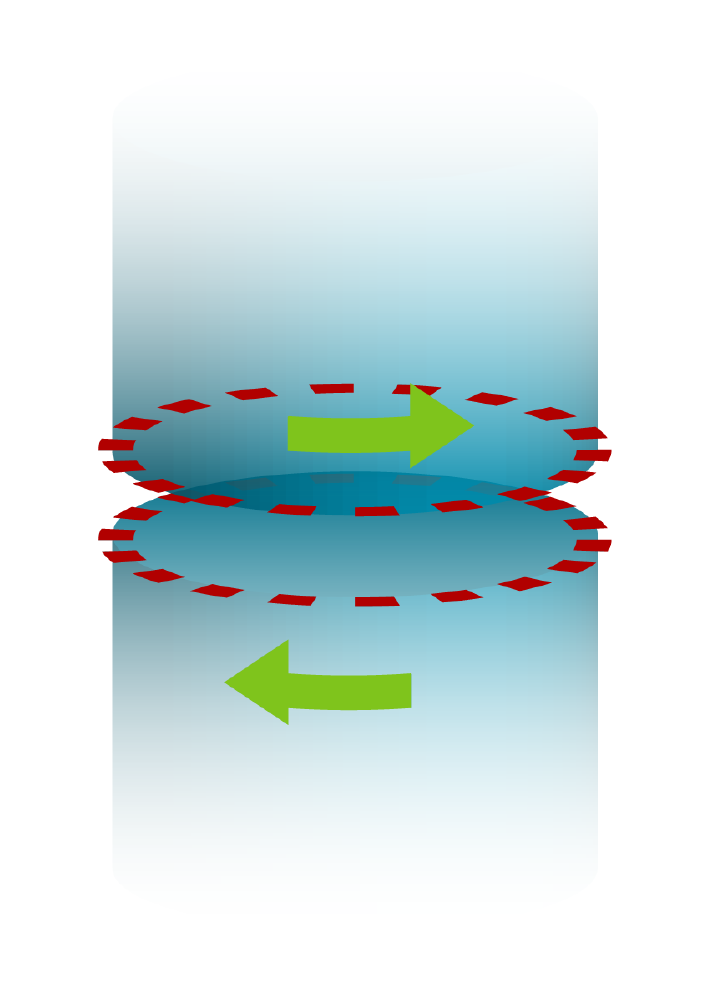}
			\caption{}\label{infinite}
		\end{subfigure}
		\begin{subfigure}[b]{.49\linewidth}
			\centering
			\includegraphics[width=0.75\linewidth,clip,bb=0 0 205 284]{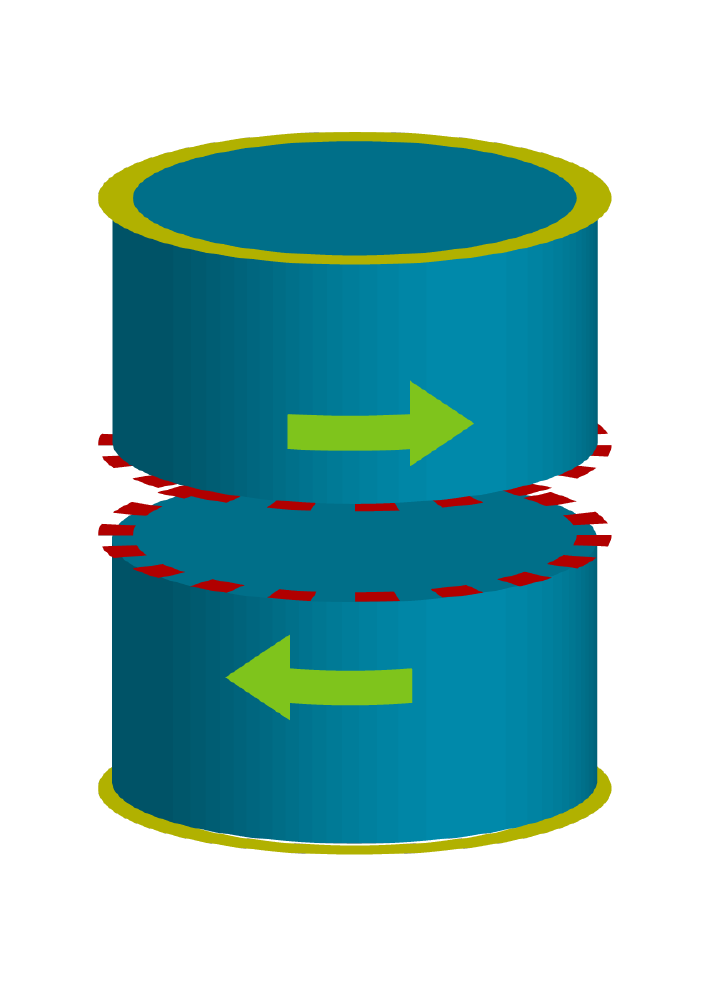}
			\caption{}\label{finite}
		\end{subfigure}
		\caption{Two types of the two-dimensional Ising model with a slip line (red broken lines): (\subref{infinite}) The infinite-size model with no boundaries; (\subref{finite}) Our finite-size model with two open boundaries (yellow solid lines), on which we impose various boundary conditions.}
		\label{size}
	\end{figure}

	We here explore behavior of the magnetic friction by considering an Ising ferromagnetic system from two new points of view, namely the effects of finite thickness and boundary conditions (see Fig.~\ref{size}). Many facts with the magnetic friction have been revealed, but almost all of them are related to the model of infinite size (Fig.~\ref{size}(\subref{infinite})) \cite{Hucht2012,Igloi2011,Hilhorst2011,Hucht2009,Li2016,Angst2012}, where almost exclusively non-equilibrium phase transitions are discussed. In order to understand the non-equilibrium nature of classical spin systems, however, finite-size extension is one of the most important directions. We have made the system size finite in the direction perpendicular to the slip line and apply two types of boundary conditions on the top and bottom of the system (Fig.~\ref{size}(\subref{finite})).
	
\begin{figure}[h]
		\begin{subfigure}[b]{.49\linewidth}
			\includegraphics[width=0.75\linewidth,clip,bb=0 0 205 284]{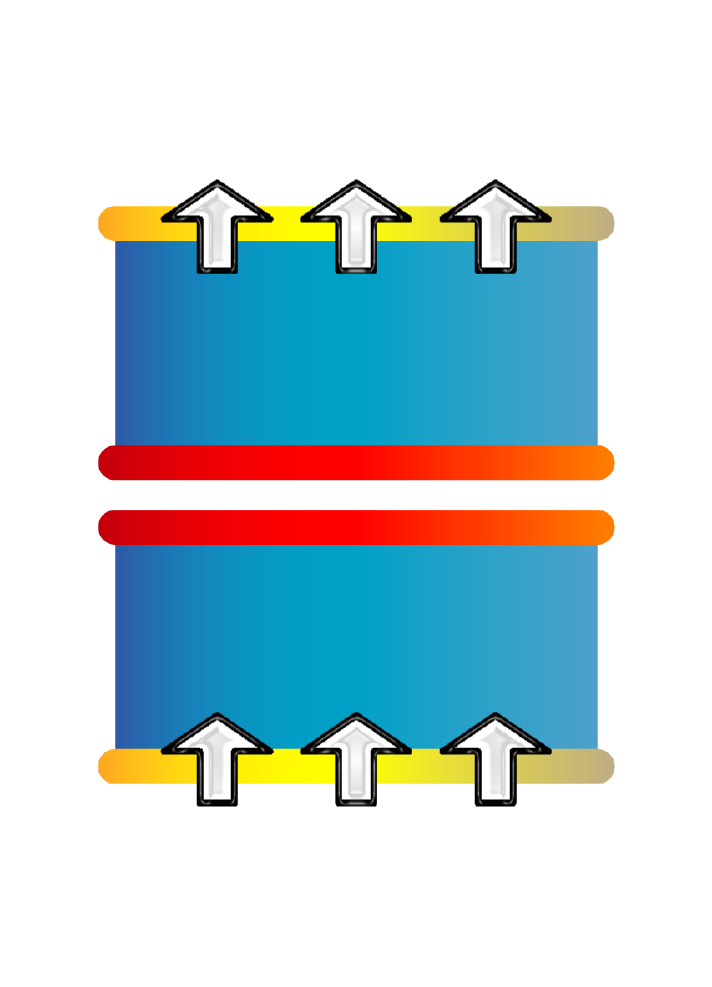}
			\caption{}\label{AP}
		\end{subfigure}
		\begin{subfigure}[b]{.49\linewidth}
			\includegraphics[width=0.75\linewidth,clip,bb=0 0 205 284]{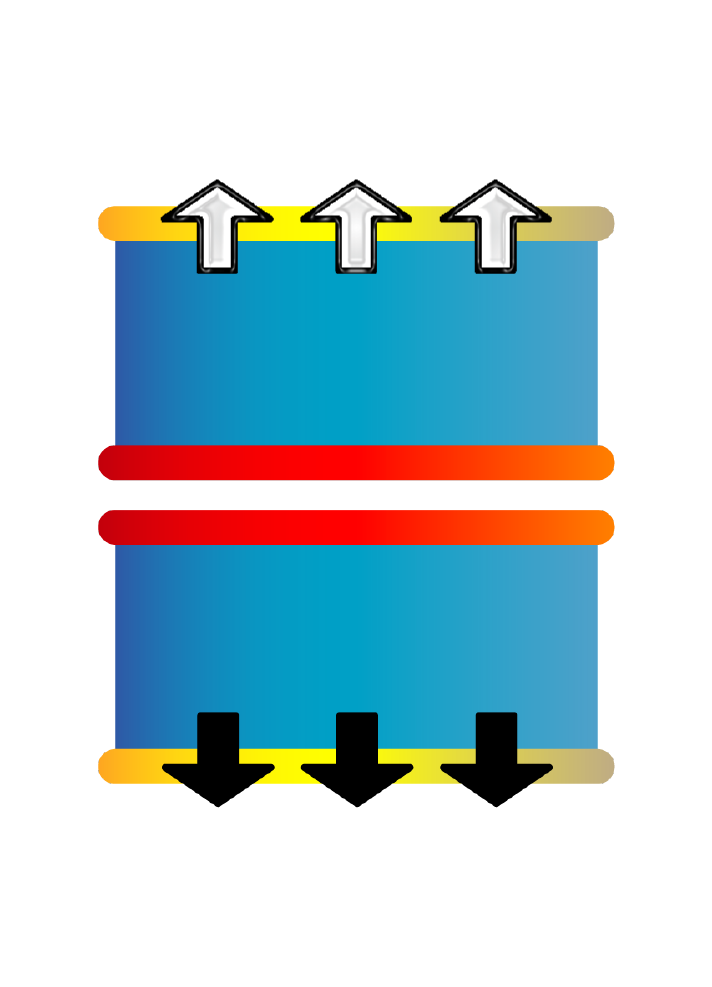}
			\caption{}\label{P}
		\end{subfigure}
		\caption{Two types of the extreme boundary condition: (\subref{AP}) the parallel boundary condition; (\subref{P}) the anti-parallel boundary condition.}
		\label{BCs}
	\end{figure}

	Our two-dimensional Ising model has two parameters, namely the temperature $T$ and the sliding velocity $v$ \cite{Kadau2008}, in addition to the sample thickness and the boundary conditions. Since we are interested in the size dependence in the direction perpendicular to the slip line ($z$ direction), we take the thermodynamic limit only in the direction parallel to the slip line ($x$ direction). This process enables us to examine how extensive physical quantities depend on the thickness $L_{z}$ in the $z$ direction. We compare them under two extreme fixed boundary conditions, namely anti-parallel and parallel conditions (see Fig.~\ref{BCs}). Because of the finite thickness, we can discuss boundary-condition dependence of physical quantities, especially the frictional force.
	
	\section{Numerical Simulations}\label{sec:Numerical}
	Our two-dimensional Ising model is described by the following Hamiltonian:
	\begin{align}
	&H=H_{\rm lower} + H_{\rm upper} + H_{\rm slip}(t),
	\end{align}
	where
	\begin{align}
	H_{\rm lower}:=&-J\sum_{x=1}^{L_{x}}\sum_{z=1}^{L_{z}/2}\sigma_{x, z}\sigma_{x+1, z}\nonumber\\
	&- J\sum_{x=1}^{L_{x}}\sum_{z=1}^{L_{z}/2-1}\sigma_{x, z}\sigma_{x, z+1}, \\
	H_{\rm upper}:=&-J\sum_{x=1}^{L_{x}}\sum_{z=L_{z}/2+1}^{L_{z}}\sigma_{x, z}\sigma_{x+1, z}\nonumber\\
	&- J\sum_{x=1}^{L_{x}}\sum_{z=L_{z}/2+1}^{L_{z}-1}\sigma_{x, z}\sigma_{x, z+1}, \\
	H_{\rm slip}(t):=&-J\sum_{x=1}^{L_{x}}\sigma_{x, L_{z}/2}\sigma_{x+vt, L_{z}/2+1}
	\end{align}
	with the ferromagnetic interaction $J>0$ and the Ising spin variables $\sigma_{x, z}=\pm 1$ ($1\leq x\leq L_{x},1\leq z\leq L_{z}$). A slip line runs at the center between the $(L_{z}/2)$th and the $(L_{z}/2+1)$th layers, along which the interactions are replaced at a constant velocity.
	
	In the case of $v=0$, our model exhibits the corresponding equilibrium state and various configurations of domain walls under a temperature $T$. If we add an external force on the system and make it slide, the domain walls around the slip line temporally break and thus the total energy rises. Subsequently, the system tries to return to equilibrium and thus the energy decreases. As the result of these two competing effects, the system exhibits a non-equilibrium steady state which depends both on the temperature $T$ and the sliding velocity $v$.
	
	\begin{figure}[h]
		\begin{subfigure}[b]{.49\linewidth}
			\includegraphics[width=1.0\linewidth,clip,bb=0 0 360 252]{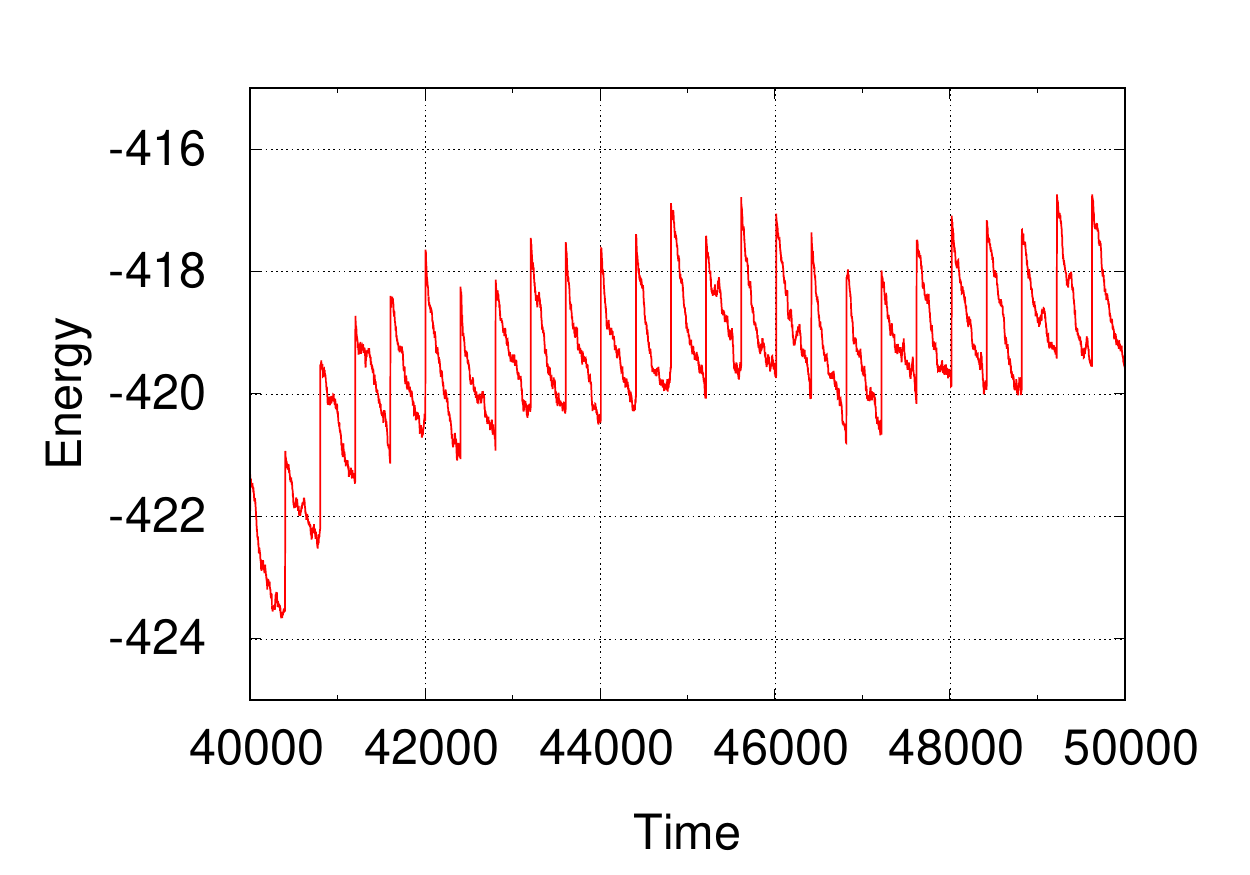}
			\caption{}\label{uness}
		\end{subfigure}
		\begin{subfigure}[b]{.49\linewidth}
			\includegraphics[width=1.0\linewidth,clip,bb=0 0 360 252]{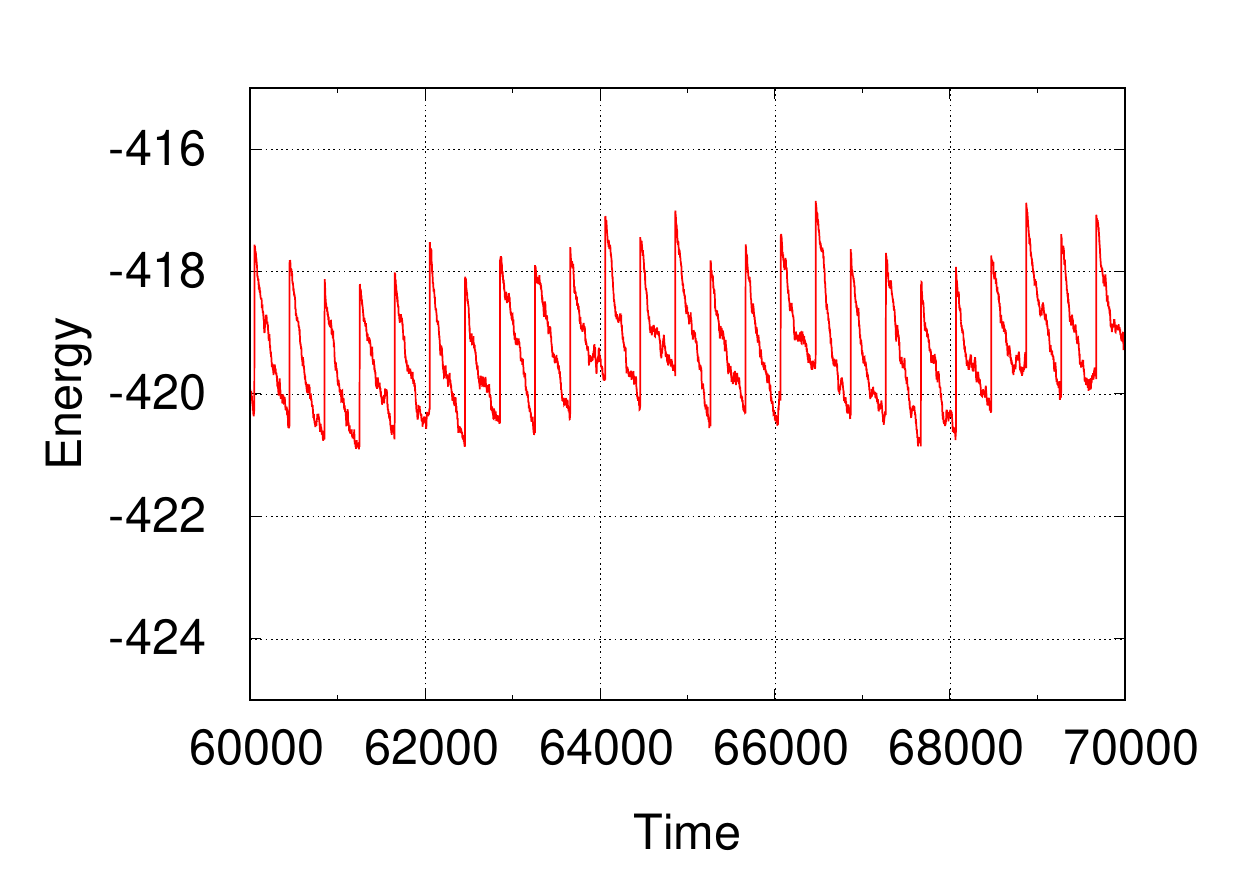}
			\caption{}\label{ness}
		\end{subfigure}
		\caption{Typical time series of the energy change against the number of single-spin flips as the Monte Carlo time: (\subref{uness}) non-equilibrium relaxation from the equilibrium state and (\subref{ness}) the non-equilibrium stationary state. We performed the simulation with the size $L_{x}\times L_{z}=20\times 20$, the temperature $T=2.5$ and the velocity $v=1$.}
		\label{timeseries}
	\end{figure}
	
	In the Monte Carlo simulation, we realize the sliding motion of the system by means of discrete sliding over a lattice constant every $1/v$ unit time and realize relaxation to the equilibrium by a sequence of single-spin flips (see Fig.~\ref{timeseries}). We implement our Monte Carlo simulations with single-spin-flip dynamics according to Ref.~\cite{Kadau2008}. The actual Monte Carlo schedule is as follows: (i) We try to relax the system to equilibrium; (ii) We slide the upper part of the system by a lattice constant against the lower; (iii). We perform single-spin flips $N/v$ times. Repeating these steps $v$ times corresponds to one Monte Carlo sweep, which therefore consists of $v$ times of slides and $N$ times of single-spin flips. 
	
	After the non-equilibrium steady state is achieved, the average energy increase due to the discrete slides per a unit time is equal to the average power done by the external force, which is the reaction to the frictional force in the steady state. Therefore, the frictional force is given by the expectation value of the energy change due to the slides over a unit time:
	\begin{align}
	F=&\left\langle \Delta H_{\rm slip}\left(t\right)\right\rangle_{\rm st}/v,
	\end{align}
	where $\Delta H_{\rm slip}$ denotes the change due to the discrete slides over a unit time and $\left\langle\bullet\right\rangle_{\rm st}$ the steady-state expectation value. We extrapolate the thermodynamic limit in the $x$ direction so that the size dependence of $F/L_{x}$ may disappear.
	
	\section{Results}\label{sec:Results}
	In the following Monte Carlo results, we took the average over 480 samples of continuous measurements over 3200 sweeps after the initial relaxation of 3200 sweeps. We adopted as the initial configuration a domain wall on the slip line in the case of the anti-parallel boundary conditions and the completely magnetized configuration in the case of parallel boundary conditions. We chose these initial states because they are the most natural ground states that are consistent with their boundary conditions.
	
\begin{figure}[h]
		\begin{subfigure}[b]{.49\linewidth}
			\includegraphics[width=.85\linewidth,clip,bb=0 0 360 252]{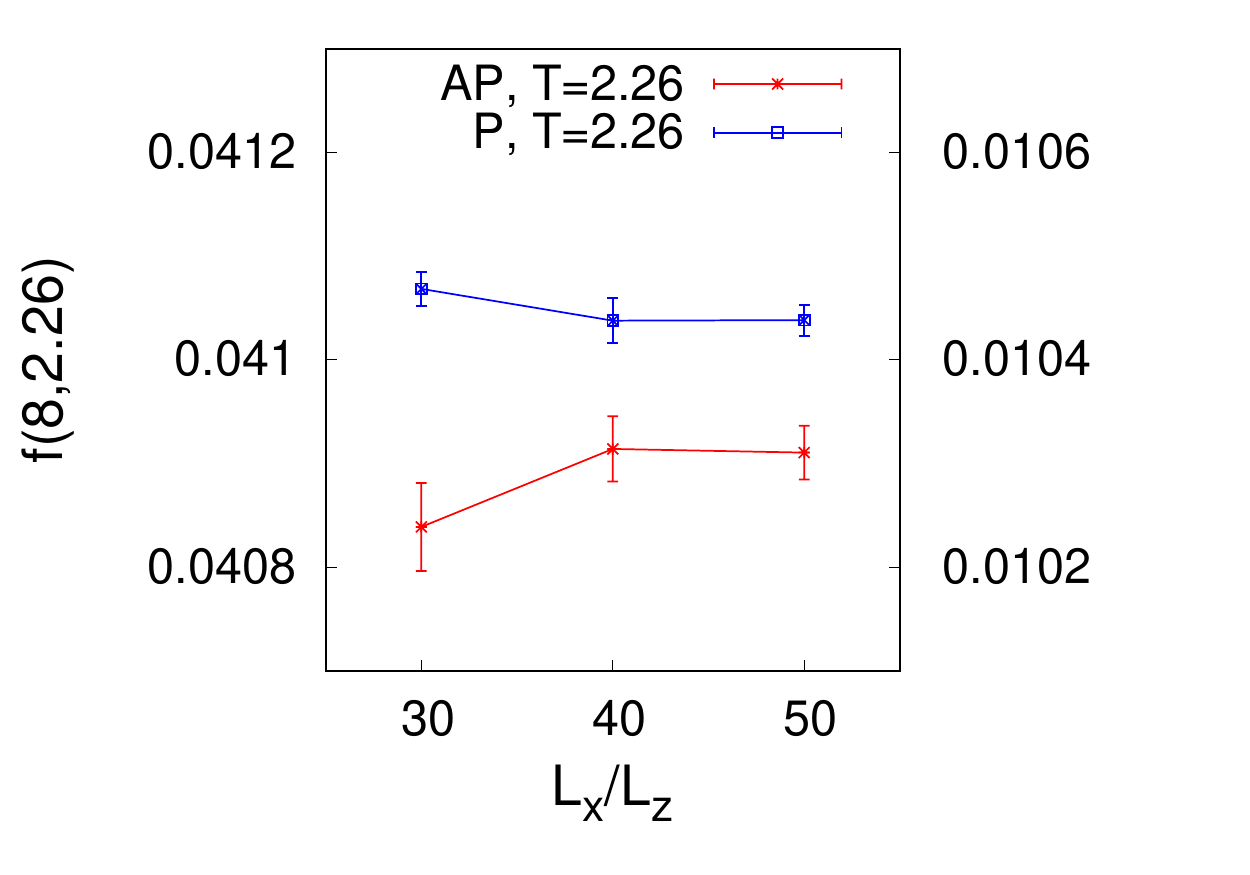}
			\caption{}\label{fricconv}
		\end{subfigure}
		\begin{subfigure}[b]{.49\linewidth}
			\includegraphics[width=0.85\linewidth,clip,bb=0 0 360 252]{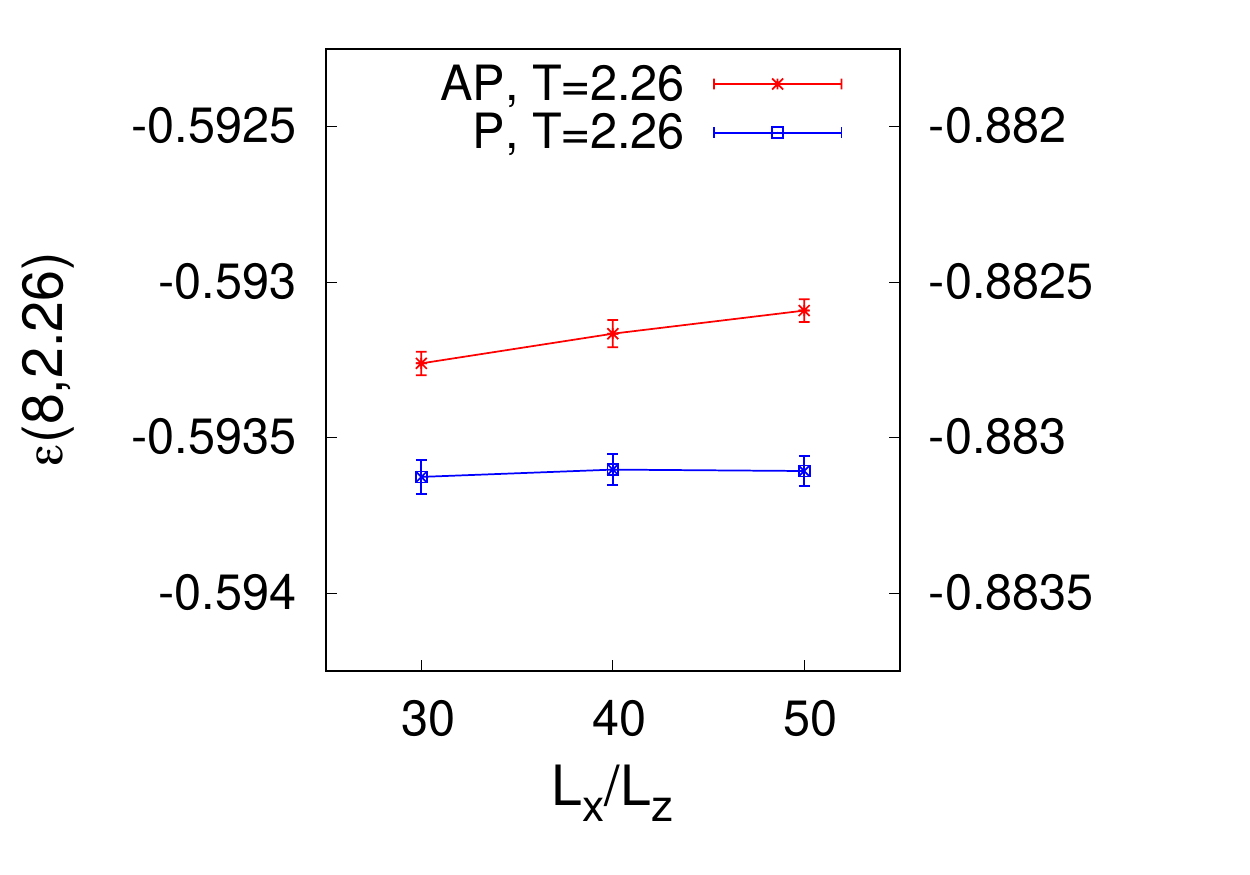}
			\caption{}\label{ebconv}
		\end{subfigure}
		\caption{The convergence in $L_{x}\to\infty$ of the physical quantities $F/L_{x}$ and $E/(L_{x}L_{z})$ for $L_{z}=8$ and $T=2.26$.}\label{fig:conv}
	\end{figure}

\begin{figure}[h]
		\centering
		\includegraphics[width=0.85\linewidth,clip,bb=0 0 360 252]{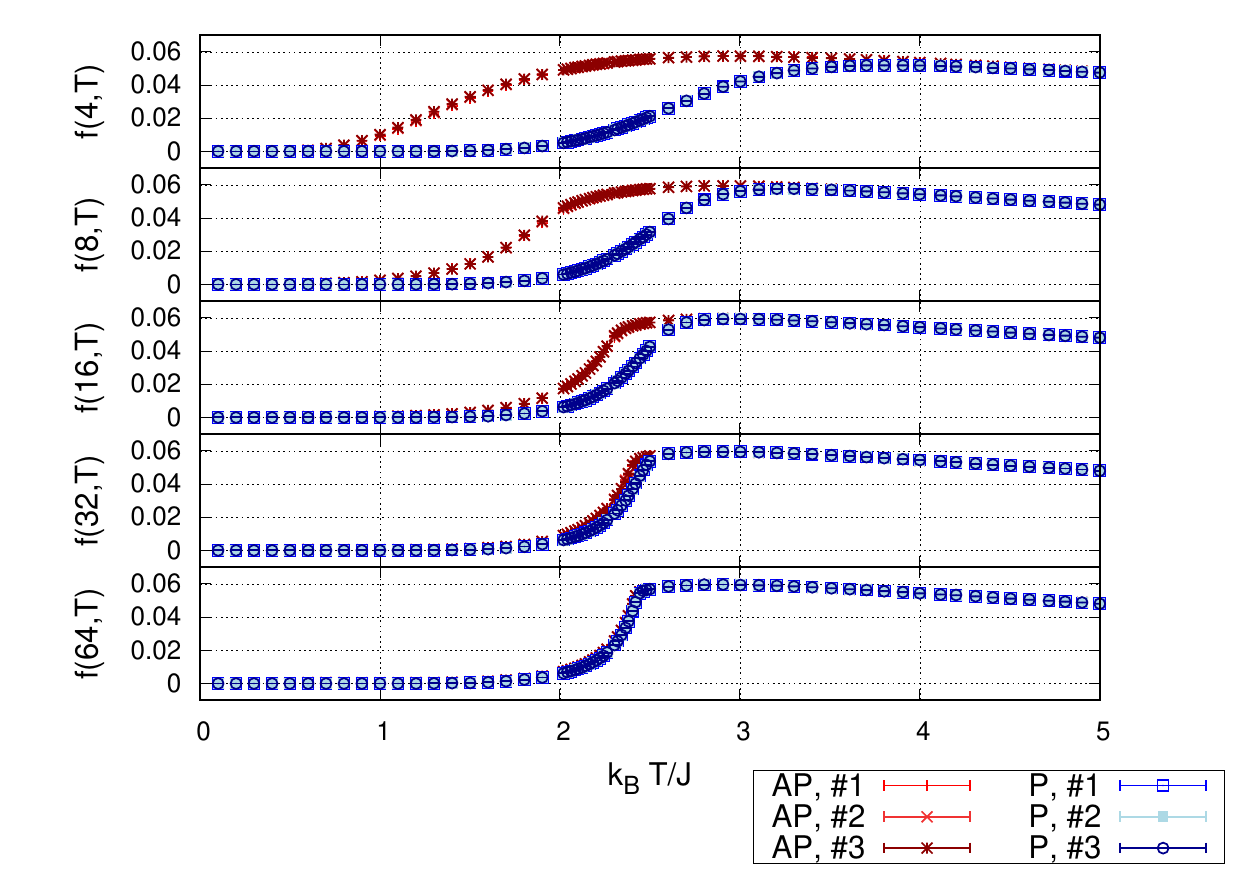}
		
		\caption{The Monte Carlo estimates of $f(L_{z},T)=F/L_{x}$ against the temperature $T$ for $L_{z} = 4, 8, 16, 32, \text{ and }64$ from the top panel to the bottom. The symbol AP and P in the key denotes the anti-parallel and the parallel boundary conditions, while the numbers \#1, \#2, and \#3 in the key correspond to the size $L_{x}/L_{z} = 30, 40, \text{ and }50$, respectively}.
		\label{fig:f}
	\end{figure}

	For each value of $L_{z}$, we estimated $F/L_{x}$ in the cases of $L_{x}=30\times L_{z},40\times L_{z},50\times L_{z}$, from which we judged that all the data well approximate the limit $L_{x}\to\infty$ within the error-bar range (see Fig.~\ref{fig:conv} (\subref{fricconv})). We find a clear difference between the estimates for the two types of boundary condition (see Fig.~\ref{fig:f}),  until it virtually vanishes for $L_{z}=64$.
	
\begin{figure}[h]
		\begin{subfigure}[b]{.49\linewidth}
			\includegraphics[width=0.75\linewidth,clip,bb=0 0 192 86]{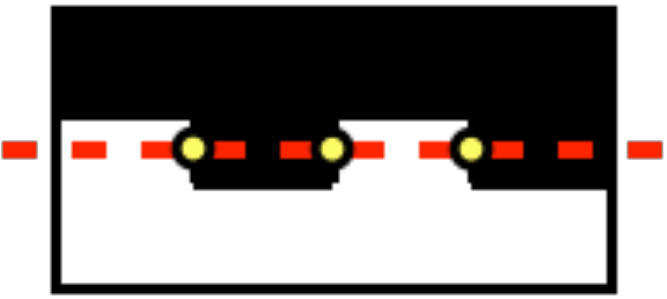}
			\caption{}\label{manycross}
		\end{subfigure}
		\begin{subfigure}[b]{.49\linewidth}
			\includegraphics[width=0.75\linewidth,clip,bb=0 0 192 86]{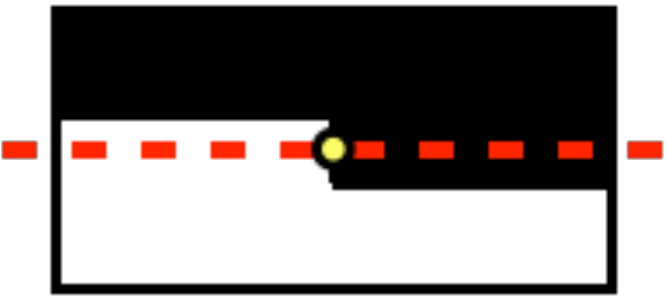}
			\caption{}\label{fewcross}
		\end{subfigure}
		\caption{The case (\subref{manycross}) with a larger spatial fluctuation has more crossing points with the slip line than the case (\subref{fewcross}) and therefore exhibits a larger frictional force.}
		\label{numcross}
	\end{figure}

	The following reasons may explain why it is achieved that the frictional force under the anti-parallel condition becomes larger than that under the parallel condition with $T\simeq T_{\rm c}$ and sufficiently small $L_{z}$: When the system is approximately two-dimensional, namely in the limit $L_{z}\to\infty$, the ratio between the correlation length and the size, $\xi_{z}(L_{z})/L_{z}$, approaches 1 around $T\simeq T_{\rm c}$. This effect is not diminished even when the size $L_{z}$ is sufficiently small as long as the condition $T\simeq T_{\rm c}$ holds, and hence the system has a strong correlation in the $z$ direction. Therefore the anti-parallel boundary condition stabilizes the domain wall along the slip line in the initial state through the long-range correlation, competing with the temperature fluctuation. The spatial fluctuation of the domain wall increases the number of crossing points between the domain wall and the slip line (Fig.~\ref{numcross}), which raises the frictional force.
	
	Our results also suggest transitions between two different steady states due to switching of boundary conditions: the frictional force can be enhanced or reduced by switching the boundary condition between the parallel and anti-parallel ones.
	
\begin{figure}[h]
		\centering
		\includegraphics[width=0.85\linewidth,clip,bb=0 0 595 417]{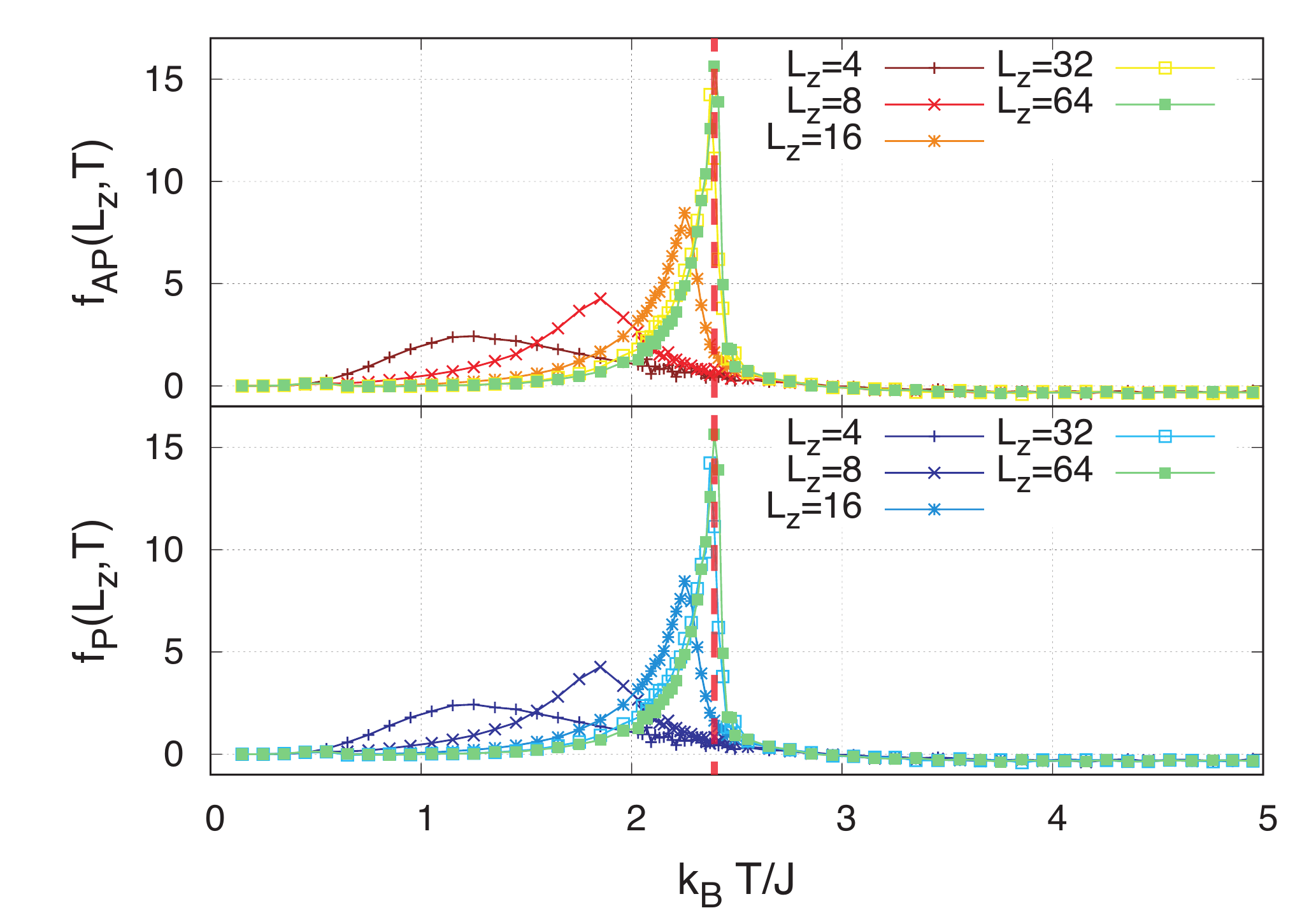}
		
		\caption{The temperature derivative $(\partial F/\partial T)/L_{x}$ against the temperature $T$ for $L_{z} = 4, 8, 16, 32, \text{ and }64$ under anti-parallel and parallel boundary conditions. The red broken vertical line indicates a non-equilibrium critical temperature $T_{\rm c}(v)|_{v=10}=2.4$, which was reported in Ref.~\cite{Hucht2009}.}
		\label{fig:df}
	\end{figure}

\begin{figure}[h]
		\centering
		\includegraphics[width=0.85\linewidth,clip,bb=0 0 360 252]{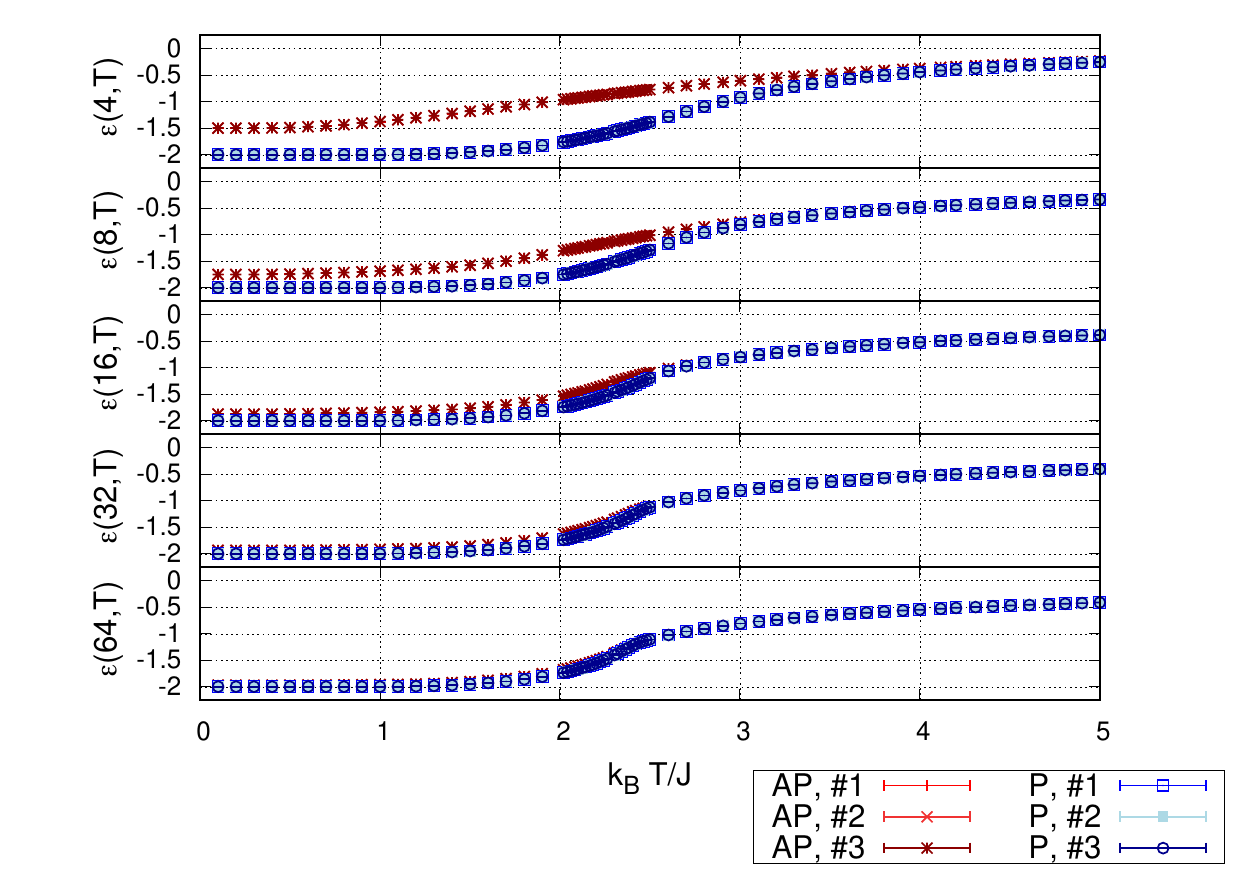}
		
		\caption{The Monte Carlo estimates of $\epsilon(L_{z},T)=E/(L_{x}L_{z})$ against the temperature $T$ for $L_{z} = 4, 8, 16, 32, \text{ and }64$ from the top panel to the bottom. The meaning of the symbols AP and P, and the numbers \#1, \#2, and \#3 is the same as Fig.~\ref{fig:f}.}
		\label{fig:eb}
	\end{figure}

\begin{figure}[h]
		\centering
		\includegraphics[width=0.85\linewidth,clip,bb=0 0 595 417]{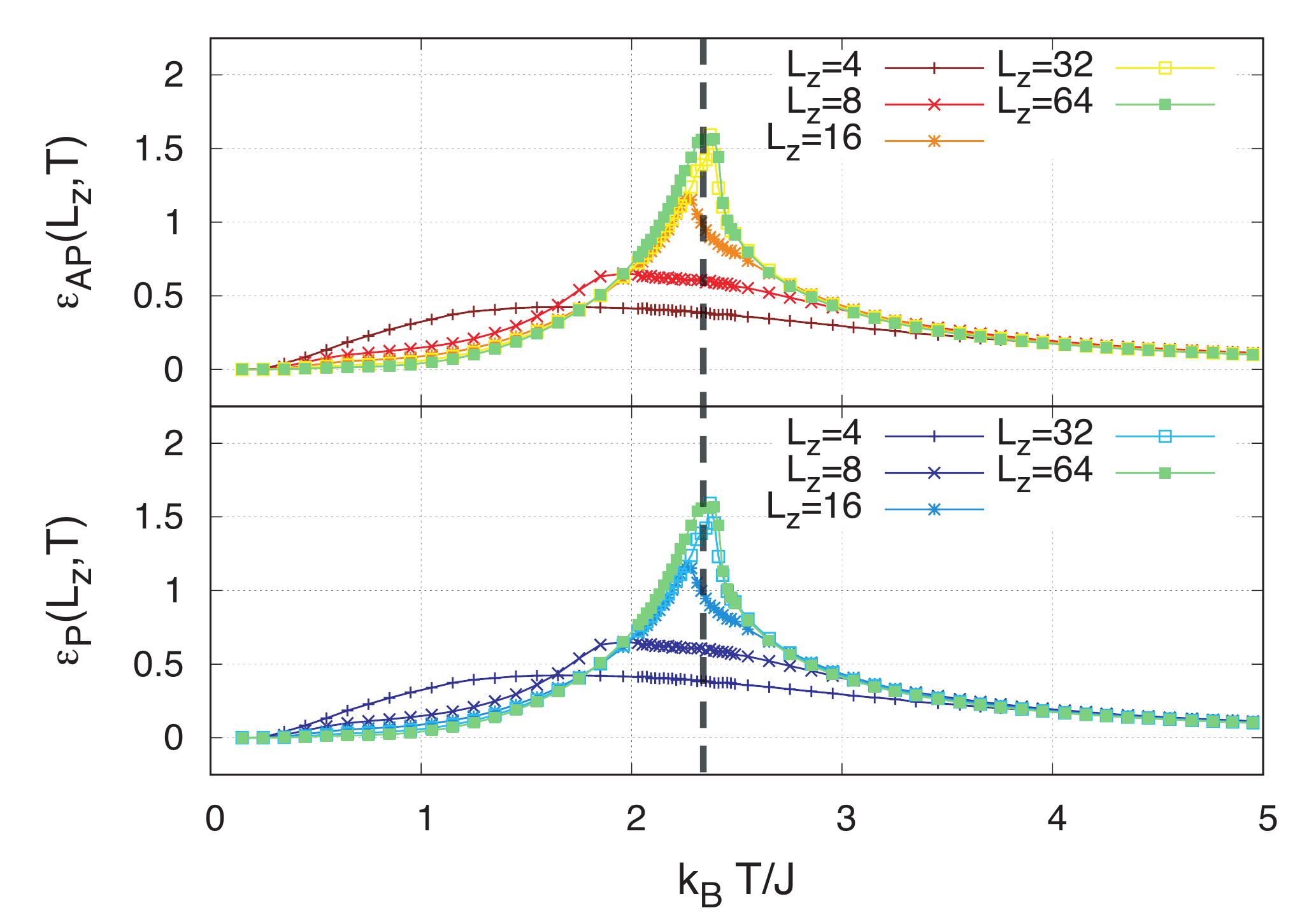}
		
		\caption{The temperature derivative $(\partial E/\partial T)/(L_{x}L_{z})$ against the temperature $T$ for $L_{z} = 4, 8, 16, 32, \text{ and }64$ under anti-parallel and parallel boundary conditions. The black-broken vertical line indicates the equilibrium critical temperature $T_{\rm c,eq}=2.26$.}
		\label{fig:deb}
	\end{figure}

	Incidentally, the temperature derivative of the frictional force indicates a peak growing for larger size $L_{z}$ (see Fig.~\ref{fig:df}). We observed that the peak location converges to $T\simeq 2.4$ in the limit $L_{z}\to\infty$, which is consistent with the non-equilibrium critical temperature $T_{\rm c,eq}(v)|_{v=10}\simeq 2.4$ reported in Ref.~\cite{Hucht2009}. We also measured the energy density of the whole system $E/(L_{x}L_{z})$ as a bulk quantity (see Fig.~\ref{fig:eb} and Fig.~\ref{fig:conv} (\subref{ebconv})). Its temperature derivative $(\partial E/\partial T)/(L_{x}L_{z})$ exhibits a peak at the temperature $T\simeq 2.26$ (see Fig.~\ref{fig:deb}), which is consistent with the exact solution of the two-dimensional Ising model.
	

	\section{Summary and Discussions}\label{sec: Summary and Discussions}
	
	We considered a two-dimensional cylindrical Ising model with a one-dimensional circular slip line, and made its size in the perpendicular direction to the slip line finite, in order to introduce non-trivial conditions on the two open boundaries. We then discussed the boundary-condition dependence of physical quantities.
	
	We found that the boundary condition imposed on the top and bottom edges of the system has a non-negligible effect on the frictional force when the thickness $L_{z}$ is small and the temperature $T$ is close to the non-equilibrium critical temperature $T_{\rm c,eq}(v)$. The frictional force under the anti-parallel condition is greater than the case of the parallel condition.
	
	A large value of the ratio $\xi_{z}/L_{z}$ of the correlation length $\xi_{z}$ along the $z$ direction and the size $L_{z}$ in the vicinity of the equilibrium critical temperature $T_{\rm c,eq}^{(2)}=2.26$ makes the difference of the frictional forces greater around the critical temperature. This effect diminishes around $L_{z}\simeq 64$ and thus the system is almost two-dimensional for $L_{z}=64$.
	
	
	We expect that our model is applicable to technical control of friction in practical situations by aligning selectively the spins on the boundaries. The magnetic contribution of friction estimated from actual magnetic metal may be comparable to other contributions, such as phonon and orbital motion of electrons \cite{Kadau2008}. According to Monte Carlo simulations, this method of control is effective when two boundaries are close enough. 
	
	As future works, we plan to discover other examples in which switching the boundary conditions for finite-thickness systems makes a remarkable difference on the frictional force, suggest an experimental method of effectively controlling the frictional force by switching the boundary conditions at the temperature with the maximum difference of frictional forces between the two boundary conditions, and make the model calculation more precise to realize such an experiment.
	
	
	\bibliographystyle{apsrev4-1}
	\bibliography{library.bib}
\end{document}